\documentstyle[12pt]{article}

\begin{document}

\title{Miller-Good Method for Symmetric Double Potential Wells}

\author{Luis Alberto del Pino\thanks{Also
at Departamento de F\'{\i}sica, 
Facultad de Monta\~na, Universidad de Pinar del Rio, Cuba}\\ 
{\it  Instituto de F\'{\i}sica, Universidad de Guanajuato} \\ 
{\it Le\'on, Guanajuato, M\'exico.}\\ 
and \\ 
Hip\'olito Mena \\
{ \it Departamento de F\'{\i}sica, Facultad de Monta\~na,}\\
{\it Universidad de Pinar del Rio, Pinar del Rio, Cuba}}

\maketitle

\begin{abstract}
The ground state energy of the quartic anharmonic oscillator is calculated 
by employing  the Miller-Good method. For this purpose an extension of the 
procedure is  developed which is  suitable   for considering four 
turning points situations. A criterion for the selection of the auxiliary 
quantum mechanical problem is also advanced.
\end{abstract}

\newpage

\section{Introduction}

 In  many  physical situations of interest, 
 quantum mechanical particles move 
under the action of potential having various minimal points. A relevant one 
of this problems is the oscillatory movement of 
the Nitrogen molecule between 
both sides of the hydrogenic triangle in the ammonia molecule. Numerous 
models and methods for their solutions have been  proposed for this        
system \cite{swa} \cite{new}
\cite{man} \cite{wall}.

   A main  objective of the present work is to illustrate a way to 
extended the Miller Good (MGM) method for the WKB studies of symmetric 
double wells by exploiting the symmetry properties \cite{MGM}.  The 
extension is 
obtained by proposing a modification of  the calculation scheme  of the 
method.  
In addition a criterion for the appropriate 
selection of the auxiliary quantum 
mechanical problem is also introduced.  The procedure is here applied 
to the calculation of the ground state energy of the quartic 
anharmonic oscillator. Within  the range of the parameters in which the 
technique is applicable the calculated values  
give  satisfactory  estimates 
of the ground state energy. In particular they 
improve the predictions of 
another approximate approach recently proposed  in 
the literature \cite{RMF},    
and gives numerical results of  similar quality  that 
the ones presented in
\cite{lanl1}.

\section{ Extension of the  Miller-Good Method }

 The calculation scheme of the Miller-Good (MGM) method  is 
 described as follows \cite{MGM}. As a solution for  \\

\begin{equation}
\psi{''}(x)+\frac{p^2(x)}{\hbar^2}\psi(x)=0
\end{equation}
where $p(x)=2m\sqrt{E-v(x)},$ the following change of variables is 
proposed \\

\begin{equation}
\psi(x)=\frac{1}{\sqrt{s{'}(x)}}\phi[s(x)]
\end{equation} 
where $\phi$ is a known solution  of the auxiliary quantum 
mechanical problem:\\

\begin{equation}
\phi{''}(s)+\frac{P^{2}(s)}{\hbar^2}\phi(s)=0 \\
\end{equation}
defined by the also known function $ P$, and the function  
$s$, which establishes the link between the  original problem and 
the auxiliary. The function $s$ in order to furnish the mentioned 
equivalence  satisfies the equation: \\

\begin{equation}
P^{2}(s)s'^{2}=p^{2}(x)+\hbar^{2}D_s(x)
\end{equation}

\begin{equation}
D_s(x)=\frac{3}{2}\frac{s^{''2}}{s^{'2}}-\frac{1}{2}\frac{s^{'''}}{s'}
\end{equation}

Up to this point, the transformation is exact. The Miller-Good 
approximation consists in disregard  the expression $D$ defined 
by (5) in the equation for $s$ given in (4). This approximation, 
then, produces the following solution for the wave-function \\

\begin{equation}
\psi_{M-G}=\frac{1}{\sqrt{[s'_{0}(x)]}}\phi[s_{0}(x)]
\end{equation} 
in which  $s_{0}$ is determined from the equation:\\

\begin{equation}
P(s_{0})s'_{0}=p(x),
\end{equation} 
the boundary conditions :

\begin{equation}
s_{0}(x_{1})=s_{01}
\end{equation} 
where  $ x_{1}$  is a turning point,  and the integral condition

\begin{equation}
\int_{s_{01}}^{s_{02}} ds_{0}P(s_{0})=\int_{x_{1}}^{x_{2}} dxp(x)
\end{equation}

The last expression constitutes the  Miller-Good quantization rule 
which coincides with the known WKB rule  if  in the quality of  the 
auxiliary problem   defining  function is  selected 
$P(s)=\hbar^2(\alpha-s^2)$  with $\alpha=2n+1 $. 

  However, a general criterion for selecting the auxiliary  problem 
in each concrete situation is lacking.  Only qualitative guides 
supported by the experience are at hand.  It is an aim   in  
this work  to introduce a modification of the approximation scheme 
just presented  allowing  to appropriately select the auxiliary 
problem in a general  way  which  adjust relevant characteristics  
of the considered task. 

   For the above mentioned purpose, let us perform in  (1)  the 
transformation  $x=x(y) $  through which this equation becomes  

\begin{equation}
\frac{\psi^{''}}{ x^{'2}}-\frac{x^{''}}{ x^{'3}}\psi'+
\frac{p^2(y)}{\hbar^2}
\psi=0
\end{equation}

Assuming now that the equation 

\begin{equation}
\phi^{''}(s)+\frac{p^{2}(s)}{\hbar^2}\phi(s)=0
\end{equation} 
have a known exact solution and that  this system have: the 
same number of  turning points which are coincident  the original 
ones,  coinciding asymptotic behavior in relevant regions, etc.. 
Then,  the  here proposed general rule consists in  to adopt 
such a problem defined by  (11) as the auxiliary one. 
    In this case , the condition  which the  function $s$ should 
obey for assuring  the equivalence of the auxiliary problem (11)  
with the original one  in the form (10)  is given by: \\

\begin{equation}
p^{2}(s)s^{'2}(y)=x^{'2}(y)p^{2}(y)+\hbar^2[D_{s}(y)-D_{s,x}(y)]
\end{equation}

\begin{equation}
D_{s,x}=\frac{3}{2}\frac{s{''}x{''}}{s'x'}-\frac{1}{2}\frac{x{'''}}{x'}
\end{equation}

 Again in the order zero approximation consisting in disregard  
the $D$ function in the equation for $s$  it follows:

\begin{equation}
p(s_{0})s_{0}^{'}(y)=x{'}(y)p(y)
\end{equation} 
and the  initial and boundary conditions 

\begin{equation}
s_{0}(y_{1})=s_{01}
\end{equation}

\begin{equation}
\int_{s_{01}}^{s_{02}} ds_{0} p(s_{0})=\int_{y_{1}}^{y_{2}} dy x^{'}(y) 
p(y)
\end{equation} 
in which $s_{01},s_{02}$ and $y_{1},y_{2}$  are the turning points 
in the variables $s$ and $y$ respectively.

The proposed transformation of the method not only allows to select  
the auxiliary problem  in an adequate manner, but also permits to 
extend the Miller-Good method to systems having four turning points. 
Let us suppose that the inverse  of the initial transformation 
$x(y)$ used for construct the auxiliary problem became even. Then, 
if the potential is also an even function  of $x$, the same  symmetry 
is guaranteed  for the function $s_{0}$  with respect to $x$. This 
property allows to select as the auxiliary problem the one 
corresponding to two isolated wells. 

\section{Ground State Energy of the Quartic Anharmonic Oscillator} 

In this ending section, the application of the technique to the 
calculation of the ground state energy of the one particle quantum 
mechanical problem defined quartic anharmonic potential 

\begin{equation}
v(x)=\lambda x^4-
\frac{m\omega ^2}{2}x^2, \, \, \,E<0 \nonumber
\end{equation}

The $x(y)$ transformation having the required characteristic 
is given by:\\

\begin{equation}
\frac{x^2}{x_{0}^{2}}=y+\frac{1}{2}
\end{equation}
\begin{equation}
x_0=\sqrt{\frac{m\omega^2}{2\lambda}}
\end{equation}

 After this transformation, the $p$ function defining the auxiliary 
problem (11) takes the form $p^2(s_0)=\hbar^2(\alpha-s_{0}^{2}),
\alpha=2n+1$  and the equation (12) becomes

\begin{equation}
(1-\xi^2)\xi^{'2}=2\frac{z^4}{\alpha^{2}\lambda^{'2}}\frac{1-
\eta^2}{1+4z\eta}+
\frac{1}{\alpha^{2}}D_{\xi}(\eta)+\frac{4z^2}{\alpha^2}D_{\eta\xi}
\end{equation}
\begin{equation}
D_{\eta\xi}=\frac{3\xi^{''}}{2z\xi'(1+4z\eta)}+\frac{3}{2(1+4z\eta)}
\end{equation}
\begin{equation}
\xi^2=\frac{s_{0}^2}{\alpha}
\end{equation}
\begin{equation}
\eta=\frac{y}{2z}
\end{equation}
\begin{equation}
\lambda'=\frac{\hbar\lambda}{m^2\omega^3}
\end{equation}

 The ground state energy  in the zero order approximation, as 
following from (16) satisfies the  two coupled equations 

\begin{equation}
E=\frac{\hbar\omega}{\lambda'}(z^2-\frac{1}{16})
\end{equation}
\begin{equation}
\frac{\pi}{2}=\sqrt{2}\frac{z^2}{\lambda'}\int_{-1}^{1} 
\sqrt{\frac{1-\eta^2} {1+4z\eta}}d\eta
\end{equation}
\begin{equation}
z<\frac{1}{4}
\end{equation}
 
In order to qualitatively evaluate  the nature  of the results of 
the method  as applied to this problem, let us note that inside one 
of the wells, an approximate solution of the equation 

\begin{equation}
\sqrt{1-\xi^2}\xi'=\sqrt{2}\frac{z^2}{\lambda'}\sqrt{\frac{1-
\eta^2}{1+4z\eta}}
\end{equation}
\begin{equation}
\xi(-1)=-1
\end{equation} 
becomes  $\xi=\eta$,  for values of $z<<\frac{1}{4}$.   After 
substituting  this solution in (20) it follows that this terms is 
order  $z^2$, a fact which indicates that performed approximation 
should be valid for small values of  $z$.

 The Figure 1 below presents the  results for the calculation  of the 
ground state energy for various values of the  parameter $ \lambda'$,  
as evaluated  from equations  (25) y (26). Also reported are the 
energy values  of the exact solution,  and the results of the 
calculation reported  in \cite{RMF} following the method proposed 
in that work.  As it can be observed the results of the treatment  
improve for small values of $z$  and remains working well 
up to values  corresponding  to ground state energies being a half of the  
deepness of the well. In Table the numerical results 
used for the graphical 
picture are given. \\ 

{\bf Fig.1}\\

\setlength{\unitlength}{0.240900pt}
\ifx\plotpoint\undefined\newsavebox{\plotpoint}\fi
\sbox{\plotpoint}{\rule[-0.200pt]{0.400pt}{0.400pt}}%
\begin{picture}(1500,900)(0,0)
\font\gnuplot=cmr10 at 10pt
\gnuplot
\sbox{\plotpoint}{\rule[-0.200pt]{0.400pt}{0.400pt}}%
\put(220.0,113.0){\rule[-0.200pt]{292.934pt}{0.400pt}}
\put(220.0,113.0){\rule[-0.200pt]{4.818pt}{0.400pt}}
\put(198,113){\makebox(0,0)[r]{0}}
\put(1416.0,113.0){\rule[-0.200pt]{4.818pt}{0.400pt}}
\put(220.0,189.0){\rule[-0.200pt]{4.818pt}{0.400pt}}
\put(198,189){\makebox(0,0)[r]{0.1}}
\put(1416.0,189.0){\rule[-0.200pt]{4.818pt}{0.400pt}}
\put(220.0,266.0){\rule[-0.200pt]{4.818pt}{0.400pt}}
\put(198,266){\makebox(0,0)[r]{0.2}}
\put(1416.0,266.0){\rule[-0.200pt]{4.818pt}{0.400pt}}
\put(220.0,342.0){\rule[-0.200pt]{4.818pt}{0.400pt}}
\put(198,342){\makebox(0,0)[r]{0.3}}
\put(1416.0,342.0){\rule[-0.200pt]{4.818pt}{0.400pt}}
\put(220.0,419.0){\rule[-0.200pt]{4.818pt}{0.400pt}}
\put(198,419){\makebox(0,0)[r]{0.4}}
\put(1416.0,419.0){\rule[-0.200pt]{4.818pt}{0.400pt}}
\put(220.0,495.0){\rule[-0.200pt]{4.818pt}{0.400pt}}
\put(198,495){\makebox(0,0)[r]{0.5}}
\put(1416.0,495.0){\rule[-0.200pt]{4.818pt}{0.400pt}}
\put(220.0,571.0){\rule[-0.200pt]{4.818pt}{0.400pt}}
\put(198,571){\makebox(0,0)[r]{0.6}}
\put(1416.0,571.0){\rule[-0.200pt]{4.818pt}{0.400pt}}
\put(220.0,648.0){\rule[-0.200pt]{4.818pt}{0.400pt}}
\put(198,648){\makebox(0,0)[r]{0.7}}
\put(1416.0,648.0){\rule[-0.200pt]{4.818pt}{0.400pt}}
\put(220.0,724.0){\rule[-0.200pt]{4.818pt}{0.400pt}}
\put(198,724){\makebox(0,0)[r]{0.8}}
\put(1416.0,724.0){\rule[-0.200pt]{4.818pt}{0.400pt}}
\put(220.0,801.0){\rule[-0.200pt]{4.818pt}{0.400pt}}
\put(198,801){\makebox(0,0)[r]{0.9}}
\put(1416.0,801.0){\rule[-0.200pt]{4.818pt}{0.400pt}}
\put(220.0,877.0){\rule[-0.200pt]{4.818pt}{0.400pt}}
\put(198,877){\makebox(0,0)[r]{1}}
\put(1416.0,877.0){\rule[-0.200pt]{4.818pt}{0.400pt}}
\put(220.0,113.0){\rule[-0.200pt]{0.400pt}{4.818pt}}
\put(220,68){\makebox(0,0){0.01}}
\put(220.0,857.0){\rule[-0.200pt]{0.400pt}{4.818pt}}
\put(355.0,113.0){\rule[-0.200pt]{0.400pt}{4.818pt}}
\put(355,68){\makebox(0,0){0.02}}
\put(355.0,857.0){\rule[-0.200pt]{0.400pt}{4.818pt}}
\put(490.0,113.0){\rule[-0.200pt]{0.400pt}{4.818pt}}
\put(490,68){\makebox(0,0){0.03}}
\put(490.0,857.0){\rule[-0.200pt]{0.400pt}{4.818pt}}
\put(625.0,113.0){\rule[-0.200pt]{0.400pt}{4.818pt}}
\put(625,68){\makebox(0,0){0.04}}
\put(625.0,857.0){\rule[-0.200pt]{0.400pt}{4.818pt}}
\put(760.0,113.0){\rule[-0.200pt]{0.400pt}{4.818pt}}
\put(760,68){\makebox(0,0){0.05}}
\put(760.0,857.0){\rule[-0.200pt]{0.400pt}{4.818pt}}
\put(896.0,113.0){\rule[-0.200pt]{0.400pt}{4.818pt}}
\put(896,68){\makebox(0,0){0.06}}
\put(896.0,857.0){\rule[-0.200pt]{0.400pt}{4.818pt}}
\put(1031.0,113.0){\rule[-0.200pt]{0.400pt}{4.818pt}}
\put(1031,68){\makebox(0,0){0.07}}
\put(1031.0,857.0){\rule[-0.200pt]{0.400pt}{4.818pt}}
\put(1166.0,113.0){\rule[-0.200pt]{0.400pt}{4.818pt}}
\put(1166,68){\makebox(0,0){0.08}}
\put(1166.0,857.0){\rule[-0.200pt]{0.400pt}{4.818pt}}
\put(1301.0,113.0){\rule[-0.200pt]{0.400pt}{4.818pt}}
\put(1301,68){\makebox(0,0){0.09}}
\put(1301.0,857.0){\rule[-0.200pt]{0.400pt}{4.818pt}}
\put(1436.0,113.0){\rule[-0.200pt]{0.400pt}{4.818pt}}
\put(1436,68){\makebox(0,0){0.1}}
\put(1436.0,857.0){\rule[-0.200pt]{0.400pt}{4.818pt}}
\put(220.0,113.0){\rule[-0.200pt]{292.934pt}{0.400pt}}
\put(1436.0,113.0){\rule[-0.200pt]{0.400pt}{184.048pt}}
\put(220.0,877.0){\rule[-0.200pt]{292.934pt}{0.400pt}}
\put(45,495){\makebox(0,0){E/Umin}}
\put(828,23){\makebox(0,0){ $\lambda'$}}
\put(220.0,113.0){\rule[-0.200pt]{0.400pt}{184.048pt}}
\put(1306,812){\makebox(0,0)[r]{Present work}}
\put(1350,812){\raisebox{-.8pt}{\makebox(0,0){$\Diamond$}}}
\put(220,791){\raisebox{-.8pt}{\makebox(0,0){$\Diamond$}}}
\put(355,708){\raisebox{-.8pt}{\makebox(0,0){$\Diamond$}}}
\put(423,668){\raisebox{-.8pt}{\makebox(0,0){$\Diamond$}}}
\put(490,626){\raisebox{-.8pt}{\makebox(0,0){$\Diamond$}}}
\put(558,587){\raisebox{-.8pt}{\makebox(0,0){$\Diamond$}}}
\put(625,549){\raisebox{-.8pt}{\makebox(0,0){$\Diamond$}}}
\put(760,471){\raisebox{-.8pt}{\makebox(0,0){$\Diamond$}}}
\put(896,397){\raisebox{-.8pt}{\makebox(0,0){$\Diamond$}}}
\put(1031,327){\raisebox{-.8pt}{\makebox(0,0){$\Diamond$}}}
\put(1098,293){\raisebox{-.8pt}{\makebox(0,0){$\Diamond$}}}
\put(1166,258){\raisebox{-.8pt}{\makebox(0,0){$\Diamond$}}}
\put(1233,228){\raisebox{-.8pt}{\makebox(0,0){$\Diamond$}}}
\put(1301,197){\raisebox{-.8pt}{\makebox(0,0){$\Diamond$}}}
\put(1436,138){\raisebox{-.8pt}{\makebox(0,0){$\Diamond$}}}
\put(1306,767){\makebox(0,0)[r]{Exact results}}
\put(1350,767){\makebox(0,0){$+$}}
\put(355,709){\makebox(0,0){$+$}}
\put(423,669){\makebox(0,0){$+$}}
\put(558,594){\makebox(0,0){$+$}}
\put(760,499){\makebox(0,0){$+$}}
\put(1098,390){\makebox(0,0){$+$}}
\put(1233,353){\makebox(0,0){$+$}}
\put(1436,301){\makebox(0,0){$+$}}
\sbox{\plotpoint}{\rule[-0.400pt]{0.800pt}{0.800pt}}%
\put(1306,722){\makebox(0,0)[r]{Reference [6]}}
\put(1350,722){\raisebox{-.8pt}{\makebox(0,0){$\Box$}}}
\put(355,839){\raisebox{-.8pt}{\makebox(0,0){$\Box$}}}
\put(423,686){\raisebox{-.8pt}{\makebox(0,0){$\Box$}}}
\put(558,541){\raisebox{-.8pt}{\makebox(0,0){$\Box$}}}
\put(760,419){\raisebox{-.8pt}{\makebox(0,0){$\Box$}}}
\put(1098,350){\raisebox{-.8pt}{\makebox(0,0){$\Box$}}}
\put(1233,319){\raisebox{-.8pt}{\makebox(0,0){$\Box$}}}
\put(1436,266){\raisebox{-.8pt}{\makebox(0,0){$\Box$}}}
\end{picture}

{ \bf Table 1}\\

\begin{tabular}{|c|c|c|c|c|}\hline  
$\lambda'$&$z^2$&$\frac{E}{\hbar\omega}$&$\frac{E}{\hbar\omega}
$(exacta)&$\frac{E}
{\hbar\omega}$[1]\\ \hline
0.01&0.0069&-5.55&-&-\\\hline
0.02&0.0138&-2.43&-2.43&-2.99\\\hline
0.025&0.0167&-1.83&-1.82&-1.88\\\hline
0.03&0.02&-1.41&-&-\\\hline
0.035&0.0233&-1.12&-1.12&-1.00\\\hline
0.04&0.0272&-0.88&-&-\\\hline
0.05&0.0345&-0.56&-0.63&-0.50\\\hline
0.06&0.0412&-0.35&-&-\\\hline
0.07&0.0492&-0.19&-&-\\\hline
0.075&0.0523&-0.13&-0.30&-0.26\\\hline
0.08&0.0568&-0.07&-&-\\\hline
0.085&0.0593&-0.03&-0.23&-0.20\\\hline
0.09&0.0610&-0.016&-&-\\\hline
0.1&0.0612&-0.013&-0.15&-0.13\\\hline
\end{tabular}

\section{Conclusions}

An extension of the Miller-Good method is proposed.  The approach 
can be applied to symmetric double  potential well problems. For 
this purpose, it is sufficient  to find the transformation allowing 
to transform the double well in two isolated wells. The numerical 
results  of the  application of the method to the quartic anharmonic 
potential  shows that the approach work well 
in regions where other method  
fails \cite{RMF}. The quasi classical character of the approximation 
indicates the possibility for that, in general, the extension 
proposed  become  useful  in the region of small values 
of the transmission 
coefficient through the barrier . This affirmation is 
supported   by the fact that the results shown 
in Table 1 are satisfactory 
ones for the small, values of $ \lambda'$  implying  
also small values for 
the transmission  coefficients.    
  It is need to remark that the calculation within the proposed approach 
can be improved by using a better value of the parameter $\alpha$ 
calculated from the exact solution of an improved auxiliary problem.
The useful  modification  consists in extending the 
auxiliary potential to be an
even function with respect to the point $y=-1/2$.
An approximated solution of this task 
have been obtained in Ref. \cite{wall}.   
\\ 

\section{Acknowledgements}
 We express our deep gratitude to our former advisor Dr.Igropulos
which introduced us to this field of  activity. We also would like to
extend this gratitude to 
the  members of the Group of Theoretical Physics of ICIMAF, in particular 
to  Drs. Alejandro Cabo, Aurora Perez, Augusto Gonzalez and Lic. David 
Oliva, for their advise and collaboration.
The support of CONACyT  and  the warm  hospitality of the Institute of 
Physics for the Guanajuato University , where this work was  finished, are 
also greatly appreciated.


\begin{thebibliography}{9}

\bibitem{swa}
J.D.Swalen and J.A.Ibers.J.Chem.Phys.{\bf36} (1961) 1914.
\bibitem{new}
R.R Newton and L.H.Thomas.J.Chem.Phys. {\bf16} (1948) 310.
\bibitem{man}
M.F.Manning.J.Chem.Phys. {\bf3} (1935) 136.
\bibitem{wall}
F.T. Wall and George Glocker.Journal of Chemical Physics {\bf5} (1937) 314.
\bibitem{MGM} F. Galindo et. al., Mecanica Cuantica, Editorial Reverte,
Madrid (1977).
\bibitem{RMF} N. Aquino, J.L. Lopez-Bonilla and M.A.Rosales, Rev. Mex. Fis. 
{\bf 40} (1994) 946.
\bibitem{lanl1} 
C.S. Park, S.Y. Lee, J.R. Kahng, S.K. Yoo, D.K. Park, C.H.Lee
and E.S. Ying, Los Alamos Preprint Archive, quant-ph/9609008 
(9 Sept. 1996).
\end{thebibliography}
\end{document}